\documentstyle[12pt,epsfig]{article}
\begin{document}
\oddsidemargin .4cm
\begin{titlepage}
\title{
\hfill {\normalsize DTP/97/50}\\
\hfill{\normalsize hep-ph/9707366}\\
\hfill{\normalsize June 1997}\\
Renormalization scheme dependence and the problem of 
 determination of $\alpha_{s}$ and the condensates from the
semileptonic $\tau$ decays.} 
\author{ P. A. R\c{a}czka\thanks{On leave of absence from Institute
of Theoretical Physics, Warsaw University, Warsaw, Poland.}\\
Centre for Particle Theory, University of Durham\\
South Road, Durham, DH1~3LE, Great Britain}
\date{\quad}
\maketitle
\begin{abstract}
The QCD corrections to the moments of the invariant mass
distribution in the semileptonic $\tau$ decays are considered.
The effect of the renormalization scheme dependence on the fitted
values of $\alpha_{s}(m^{2}_{\tau})$ and the condensates is
discussed, using a simplified approach where the nonperturbative
contributions are approximated by the dimension six condensates.
The fits in the vector and the axial-vector channel are
investigated in the next-to-leading and the
next-to-next-to-leading order. The next-to-next-to-leading order
results are found to be relatively stable with respect to change
of the renormalization scheme. A change from the
$\overline{\mbox{MS}}$ scheme to the minimal sensitivity scheme
results in the reduction of the extracted value of
$\alpha_{s}(m^{2}_{\tau})$ by 0.01.

\begin{center}
\end{center}

\end{abstract}
\thispagestyle{empty}
\end{titlepage}
\setcounter{page}{1} 
\newpage

\section{Introduction}

Recently there has been considerable progress in the
determination of the vector and axial-vector hadronic spectral
functions from the semileptonic decays of the $\tau$ lepton
\cite{aleph93}--\cite{aleph97}. Using the
QCD predictions for the moments of these 
spectral functions one may obtain constraints on $\alpha_{s}$ and
the parameters characterizing the nonperturbative QCD dynamics
(\cite{bra88}--\cite{nari93}, \cite{domi88}--\cite{kart92}). The
accuracy of the obtained 
value of $\alpha_{s}$ 
appears to be quite high.
In order to have a proper understanding of the phenomenological
relevance of this determination of $\alpha_{s}$ it is important
to make a careful estimate the theoretical uncertainties in the
QCD predictions \cite{pich95}--\cite{milt97}. In this note we
investigate the 
uncertainties in the evaluation of the perturbative part of the
QCD prediction for the spectral moments, for which the
next-to-next-to-leading order (NNLO) approximation is available.
We study in some detail the sensitivity of the perturbative QCD
predictions
to the choice of the renormalization scheme (RS) and the effect of the
RS dependence on the fitted values of the QCD parameters.

\section{Theoretical framework}

Let us begin by summarizing the theoretical framework adopted in the
analysis of the $\tau$ decay data \cite{bra88}--\cite{nari93}. 
In \cite{ledi92b} it was suggested to use the $R^{kl}_{\tau,V/A}$
moments, defined by the relation
\begin{equation}
R^{kl}_{\tau,V/A}=\frac{1}{\Gamma_{e}} \int_{0}^{m_{\tau}^{2}}
ds\,\left(1-\frac{s}{m_{\tau}^{2}}\right)^{k}
\left(\frac{s}{m_{\tau}^{2}}\right)^{l}
\frac{d\Gamma_{ud}^{V/A}}{ds},
\label{eq:rkldef}
\end{equation}
where $d\Gamma_{ud}^{V/A}/ds$ denotes the invariant mass
distribution for the Cabbibo allowed semileptonic $\tau$ decays
in the vector (V) or axial-vector (A) channel and $\Gamma_{e}$
denotes the electronic width of the $\tau$ lepton. The QCD
predictions for $R^{kl}_{\tau,V/A}$ have the form:
\begin{equation}
R_{\tau,V/A}^{kl}=\frac{3}{2}\,|V_{ud}|^{2}\,S_{EW}\,
R^{kl}_{0}(1+\delta^{kl}_{pt}+\delta^{kl}_{npt,V/A}),
\label{eq:rklpred}
\end{equation}
where $V_{ud}$ is the CKM matrix element ($|V_{ud}|=0.9752$) and
$S_{EW}=1.0194$ represents the correction from electroweak
interactions \cite{marc88,bra90}. The $R^{kl}_{0}$ factor denotes the
parton model prediction --- for later use we shall need
$R^{00}_{0}=1$ and $R^{12}_{0}=13/210$. The $\delta^{kl}_{pt}$
term denotes the perturbative contribution, evaluated for 3
massless quarks. (The $u$ and $d$ quark mass effects are
negligible, the $s$ quark mass effects are very small for
non-strange decays  and the $c$ quark is considered to
be decoupled \cite{chet93,lari95}.) The $\delta^{kl}_{pt}$ term
is universal 
for the V and A channels. The $\delta^{kl}_{npt,V/A}$ term
in~(\ref{eq:rklpred}) denotes the contribution from the
nonperturbative QCD effects, which are estimated using the SVZ
approach~\cite{shif79}:
\begin{equation}
\delta^{kl}_{npt,V/A}=\sum_{D=4,6...}\delta^{kl}_{(D)V/A}
=\sum_{D=4,6...}\frac{1}{m^{D}_{\tau}}
\sum_{j} c^{kl}_{D,j}<O_{D,j}^{V/A}>,
\label{eq:svz}
\end{equation}
where $<O_{D,j}^{V/A}>$ are the vacuum expectation values of the
gauge invariant operators of dimension $D$ and $c^{kl}_{D,j}$ are
coefficients specific for the considered spectral moment and the
type of the operator. The $c^{kl}_{D,j}$ coefficients are in
principle power series in the strong coupling constant.

The object of greatest interest is of course the total decay rate
for the Cabbibo allowed semileptonic $\tau$ decays in the
vector/axial-vector channels, $R^{00}_{\tau,V/A}$. The
perturbative correction to this moment is sizeable and it is
highly sensitive to the value of the strong coupling,
due to the low characteristic energy scale of
$m_{\tau}=1.777~\mbox{GeV}$. The higher $R^{kl}_{\tau}$ moments
are introduced to take advantage of the the full information
contained in the hadronic spectral functions. By using the
predictions for several $R^{kl}_{\tau}$ moments one may obtain a
simultaneous fit of $\alpha_{s}(m^{2}_{\tau})$ and of some of the
condensates $<O_{D,j}^{V/A}>$. In this way the whole analysis
becomes self-consistent and in addition one obtains a check on
both the perturbative and the nonperturbative QCD contributions.

The perturbative QCD corrections $\delta_{pt}^{kl}$ are evaluated
using a contour integral expression \cite{lam77,schi84}, which
relates them to the 
QCD correction $\delta_{\Pi}^{pt}$ to the so called Adler
function \cite{adle74}, i.e. the logarithmic derivative of the
transverse part of the vector/axial-vector current correlator
$\Pi_{ud}^{(1)V/A}$:
\begin{equation}
-12\pi^{2}\,\sigma \frac{d\,}{d\sigma}
\Pi_{ud,pt}^{(1)V/A}(\sigma)=3\,[1+\delta_{\Pi}^{pt}(-\sigma)].
\label{eq:defadler}
\end{equation}
(In the approximation of massless quarks the perturbative
contributions to the Adler
functions for the vector and axial-vector current correlators are
identical.) We have: 
\begin{equation}
\delta_{pt}^{kl}=\frac{i}{\pi}\int_{C}\frac{d\sigma}{\sigma}
f^{kl}\left(\frac{\sigma}{m_{\tau}^{2}}\right)\delta_{\Pi}^{pt}(-\sigma),
\label{eq:cont}
\end{equation}
where $f^{kl}(\sigma/m_{\tau}^{2})$ is a weight function specific
to the considered moment and $C$ is a contour running clockwise
from $\sigma=m^{2}_{\tau}-i{\epsilon}$ to
$\sigma=m^{2}_{\tau}+i{\epsilon}$ away from the region of small
$|\sigma|$. In the following we shall need the weight functions
$f^{00}$ and $f^{12}$:
\begin{equation}
f^{00}(x)=\frac{1}{2}-x+x^{3}-\frac{1}{2}\,x^{4}.
\label{eq:weight00}
\end{equation}
\begin{equation}
f^{12}(x)=\frac{1}{2}-\frac{70}{13}x^{3}+
\frac{105}{26}x^{4}+\frac{126}{13}x^{5}-
\frac{175}{13}x^{6}+\frac{60}{13}x^{7}.
\label{eq:weight12}
\end{equation}

The NNLO renormalization group improved perturbative expansion
for $\delta_{\Pi}^{pt}$ may be written in the form:
\begin{equation}
\delta_{\Pi}^{pt}(-\sigma) = a(-\sigma)[1+
r_{1}a(-\sigma)+r_{2}a^{2}(-\sigma)],
\label{eq:dpert}
\end{equation}
where $a=\alpha_{s}/\pi=g^{2}/(4 \pi^{2})$ denotes the running
coupling constant that satisfies the NNLO renormalization group
(RG) equation:
\begin{equation}
\sigma \frac{d\,a}{d\sigma} = - \frac{b}{2}
\,a^{2}\,(1 + c_{1}a + c_{2}a^{2}\,).
\label{eq:rge}
\end{equation}
In the modified minimal subtraction $\overline{\mbox{MS}}$ scheme
(i.e. using $\overline{\mbox{MS}}$ subtraction procedure and
choosing $\mu^{2}=-\sigma$) we have
\cite{chet79}--\cite{gori91} for $n_{f}=3$ 
$r_{1}^{\overline{MS}}=1.63982$ and
$r_{2}^{\overline{MS}}=6.37101$. The renormalization group
coefficients  for $n_{f}=3$ are $b=4.5$, $c_{1}=16/9$
and $c_{2}^{\overline{MS}}=3863/864\approx4.471$.

The QCD predictions for the $\delta_{pt}^{kl}$ are usually
calculated in the modified minimal subtraction
($\overline{\mbox{MS}}$) renormalization scheme \cite{bard78}. However,
in the NNLO approximation with massless quarks there is a
two-parameter freedom in choosing the RS. This is a consequence
of the fact that in each order of perturbation expansion the
finite parts of the renormalization constant for the coupling
constant may be chosen arbitrarily. Different choices of the
finite parts of the renormalization constant result in different
definitions of the coupling constant, which are related by a
finite renormalization. This results in a change of values of the
coefficients $r_{1}$, $r_{2}$ and $c_{2}$. (We restrict our
discussion to the class of mass and gauge independent schemes,
for which the coefficients $b$ and $c_{1}$ are universal.) The
formulas describing how the redefinition of the coupling affects
the coefficients $r_{i}$ and $c_{2}$ are collected for example
in~\cite{racz92}. The dimensional QCD parameter $\Lambda$
also depends on the choice of the RS~\cite{celm79}. In the NNLO there
exists however a RS invariant combination of the expansion
coefficients~\cite{stev81}--\cite{grun84}:
\begin{equation}
\rho_{2}=c_{2}+r_{2}-c_{1}r_{1}-r_{1}^{2}.
\label{eq:rho2}
\end{equation}
For the $\delta_{\Pi}$ we have $\rho_{2}=5.23783$.

The change in the expansion coefficients and the change in the
coupling constant compensate each other, but of course in the
finite order of perturbation expansion such compensation may only
be approximate, which results in the numerical RS dependence of
the perturbative predictions. This RS dependence is formally of
higher order in the coupling constant, but in the case of the
$\tau$ decay it may be significant numerically, since the
coupling is not very small at the energy scale of $m_{\tau}$. It
is therefore very important to verify to what extent the RS
dependence affects the predictions and the fits to the
experimental data.

The authors of \cite{aleph93,cleo95,hoeck96} used the
$R^{kl}_{\tau}$ moments with 
$k=1$ and $l=0,1,2,3$, and fitted the $D=4,6,8$ condensates.
Since the aim of this note is primarily to study the theoretical
uncertainties in the whole procedure we shall adopt a simplified
approach, which still has considerable phenomenological
relevance. This approach is based on the fact that the dominant
nonperturbative contribution to $R^{00}_{\tau,V/A}$ in the SVZ
expansion comes from the $D=6$ term \cite{bra88,nari88},
because the $D=4$ term is suppressed by additional power of
$\alpha_{s}$. We shall therefore neglect the $D=4$ contribution
to the total decay rate and --- for consistency --- the higher
order correction to the $D=6$ coefficient. (Such an approximation
was in fact made already in \cite{nari88}.) In the following we
shall also neglect contributions from the $D\geq 8$ condensates.
In order to be able to fit the $D=6$ condensate together with
$\alpha_{s}$ we shall use the QCD prediction for the
$R^{12}_{\tau,V/A}$ moment, which similarly to
$R^{00}_{\tau,V/A}$ has a suppressed contribution from the $D=4$
condensate. Neglecting in $R^{12}_{\tau,V/A}$ the contributions
from $D\geq 8$ condensates we obtain a simple set of
self-consistent formulas. In our approximation:
\begin{eqnarray}
\delta^{00}_{npt,V/A}&=&\delta^{00}_{(6)V/A}=
-\frac{24\pi^{2}}{m_{\tau}^{6}}
\sum_{j}C_{6,j}<O_{6,j}^{V/A}>,\\
\delta^{12}_{npt,V/A}&=&\delta^{12}_{(6)V/A}=
\frac{1680\pi^{2}}{13\,m_{\tau}^{6}}
\sum_{j}C_{6,j}<O_{6,j}^{V/A}>,
\label{eq:nptcoeff}
\end{eqnarray} 
where $\sum_{j}C_{6,j}<O_{6,j}^{V/A}>$ is the leading $D=6$
contribution to the transverse part of the hadronic vacuum
polarization function:
\begin{equation}
(-\sigma)^{3}\Pi^{(1)V/A}_{ud(D=6)}(\sigma)=\sum_{j}C_{6,j}<O_{6,j}^{V/A}>.
\end{equation}
This contribution is dominated by the four-quark condensates
\cite{bra92}. In the phenomenological analysis it is usually
expressed in a simplified form, motivated by the chiral symmetry
and the vacuum saturation approximation~\cite{bra92}:
\begin{equation}
\sum_{j}C_{6,j}<O_{6,j}^{V/A}>=
h_{V/A}\frac{32\pi}{81}\alpha_{s}\rho<\bar{q}q>^{2},
\label{eq:vacsat}
\end{equation}
where $<\bar{q}q>$ is the quark condensate, $h_{V}=-7$,
$h_{A}=11$ and $\rho$ is an effective parameter, characterizing
the deviation from the strict vacuum saturation.

In our study of the RS dependence effects we parametrize the
freedom of choice of the RS by the parameters $r_{1}$ and
$c_{2}$, following the conventions of our previous
work~\cite{racz96a,racz96b,racz96c}. To obtain the
perturbative QCD corrections 
$\delta_{pt}^{kl}$ we evaluate the contour integral numerically,
using under the integral a numerical solution of the RG equation
(\ref{eq:rge}) in the complex energy plane. In this way we take full
advantage of the RG-invariance properties of the perturbative
prediction and we resum to all orders some of the large terms
which would otherwise appear in the perturbation
expansion~\cite{pivo92,ledi92a}. We assume that the integration
contour $C$ is a circle $\sigma=-m_{\tau}^{2}\exp(-i\theta)$, $\theta
\in  [-\pi,\pi]$. 
 To determine the
numerical value of the running coupling constant on the contour
$C$ we solve the transcendental equation, which results from
integration of the RG equation (\ref{eq:rge}) with a suitable
boundary condition and analytic continuation to the complex
energy plane:
\begin{equation}
b\ln\left(\frac{m_{\tau}}
{\Lambda^{(3)}_{\overline{MS}}}\right)-
i\frac{b\theta}{2}=r^{\overline{MS}}_{1}-r_{1}+
c_{1}\ln\left(\frac{b}{2c_{1}}\right)+F^{(n)}(a),
\label{eq:rgesol}
\end{equation}
where in NLO 
\begin{equation}
F^{(1)}(a)=\frac{1}{a}+c_{1}\ln\left(\frac{c_{1}a}{1+c_{1}a}\right),
\end{equation}
and in NNLO for $4c_{2}-c_{1}^{2}>0$

\begin{eqnarray}
F^{(2)}(a)&=&\frac{1}{a}+c_{1}\ln(c_{1}a)-
\frac{c_{1}}{2}\ln(1+c_{1}a+c_{2}a^{2})+\nonumber\\
&&\frac{2c_{2}-c_{1}^{2}}{(4c_{2}-c_{1}^{2})^{1/2}}\arctan\left(
\frac{a(4c_{2}-c_{1}^{2})^{1/2}}{2+c_{1}a}\right).
\end{eqnarray}
The presence of $\Lambda^{(3)}_{\overline{MS}}$ in the expression
valid for arbitrary scheme follows from our taking into account
explicitly the relation between $\Lambda$ parameters in different
schemes \cite{celm79} and using $\Lambda^{(3)}_{\overline{MS}}$ as a
reference parameter.

After evaluating the predictions for $\delta^{00}_{pt}$ and
$\delta^{12}_{pt}$ we perform the fits to the experimental data
for $R^{00}_{\tau,V/A}$ and
$D^{12}_{\tau,V/A}=R^{12}_{\tau,V/A}/R^{00}_{\tau,V/A}$. We use
the experimental values reported recently by ALEPH \cite{hoeck96}:
$R^{00}_{\tau,V}=1.782\pm0.018$,
$D^{12}_{\tau,V}=0.0532\pm0.0007$,
$R^{00}_{\tau,A}=1.711\pm0.019$,
$D^{12}_{\tau,A}=0.0639\pm0.0005$. In the fits we assume for
simplicity that the experimental errors for $R^{00}_{\tau}$ and
$D^{12}_{\tau}$ are not correlated.

We express the results of our fits in terms of
$\alpha_{s}(m_{\tau}^{2})$ and $\delta^{00}_{(6)}$. (We actually
fit the value of $\Lambda^{(3)}_{\overline{MS}}$, which we then
convert to $\alpha_{s}(m_{\tau}^{2})$ using the RG equation in
the ${\overline{\mbox{MS}}}$ scheme.) For comparison with other
determinations of the strong coupling constant we extrapolate
$\alpha_{s}$ from $m_{\tau}^{2}$ to $m_{Z}^{2}$. Our procedure
for extrapolation relies on the matching formula relating
$\alpha_{s}(\mu^{2},n_{f}+1)$ to $\alpha_{s}(\mu^{2},n_{f})$:
\begin{equation}
\alpha_{s}(\mu^{2},n_{f}+1)=\alpha_{s}(\mu^{2},n_{f})+
\frac{L}{3}\frac{\alpha_{s}^{2}(\mu^{2},n_{f})}{\pi}+
\frac{1}{9}\left(L^{2}+\frac{57}{4}L-\frac{11}{8}\right)
\frac{\alpha_{s}^{3}(\mu^{2},n_{f})}{\pi^{2}},
\label{eq:match}
\end{equation}
where $L=\ln(\mu/\tilde{m_{q}})$ and $\tilde{m_{q}}$ is the
running quark mass $m_{q}(\mu^{2})$ of the heavy quark evaluated
at the scale $\mu=m_{q}$. (The NNLO matching formula of that form
was originally proposed in \cite{wetz82,bern82,bern83}, see also
discussion in \cite{rodri93}. However, in \cite{lari95} it
was found that a numerical coefficient in the NNLO term is
actually different, which was subsequently confirmed in
\cite{chet97}. We use the coefficient of \cite{lari95}.) In order to
evolve $\alpha_{s}$ from scale $\mu_{1}$ to the scale $\mu_{2}$
we solve the equation:
\begin{equation}
F^{(k)}(\alpha_{s}(\mu^{2}_{1})/\pi)-
F^{(k)}(\alpha_{s}(\mu^{2}_{2})/\pi)=b\ln(\mu_{1}/\mu_{2}).
\label{eq:evol}
\end{equation}
To obtain $\alpha_{s}(m^{2}_{Z})$ from the given value of
$\alpha_{s}(m^{2}_{\tau})$ we first evolve $\alpha_{s}$ from the
scale $m_{\tau}$ to the scale of $2\tilde{m_{c}}$ using the
$n_{f}=3$ RG equation, then we use the matching formula
(\ref{eq:match}) to obtain $\alpha_{s}((2\tilde{m_{c}})^{2},n_{f}=4)$,
evolve this to the scale of $2\tilde{m_{b}}$ using the $n_{f}=4$
RG equation, use the matching formula to obtain
$\alpha_{s}((2\tilde{m_{b}})^{2},n_{f}=5)$, and finally evolve
this to the scale of $m_{Z}$ using the $n_{f}=5$ RG equation. We
use $\tilde{m_{c}}=1.3\,\mbox{GeV}$ and
$\tilde{m_{b}}=4.3\,\mbox{GeV}$, which are the central values
recommended by the Review of Particle Properties \cite{pdg96}. In
Table~\ref{tb:match} we give for reference some values of
$\alpha_{s}(m^{2}_{\tau})$ and $\alpha_{s}(m^{2}_{Z})$ as a
function of $\ln(m_{\tau}/\Lambda^{(3)}_{\overline{MS}})$ in the NNLO
approximation.

\begin{table}
\begin{center}
\begin{tabular}{||c||c|c|c||}
\hline
$\ln(m_{\tau}/\Lambda^{(3)}_{\overline{MS}})$ & 
 $\Lambda^{(3)}_{\overline{MS}}$&
$\alpha_{s}(m_{\tau}^{2})$ & 
$\alpha_{s}(m_{Z}^{2})$  \\
\hline
1.30 &   484 & 0.392 &   0.1261 \\

1.35 &   461 & 0.378 &   0.1249 \\

1.40 &   438 & 0.366 &   0.1236 \\

1.45 &   417 & 0.354 &   0.1224 \\

1.50 &   397 & 0.343 &   0.1212 \\

1.55 &   377 & 0.333 &   0.1201 \\

1.60 &   359 & 0.323 &   0.1190 \\

1.65 &   341 & 0.314 &   0.1179 \\

1.70 &   325 & 0.306 &   0.1168 \\

1.75 &   309 & 0.298 &   0.1157 \\

1.80 &   294 & 0.291 &   0.1147 \\

1.85 &   279 & 0.284 &   0.1136 \\

1.90 &   266 & 0.277 &   0.1126 \\
\hline
\end{tabular}   
\end{center}
\caption{Table of values of $\alpha_{s}(m^{2}_{\tau})$ and
$\alpha_{s}(m^{2}_{Z})$ in NNLO related by the matching procedure
described in the text. }
\label{tb:match}
\end{table}

\section{Fits in the vector channel}

Let us begin with the calculation in the $\overline{\mbox{MS}}$
scheme, to see how our approximate treatment compares with a more
complete analysis reported in \cite{hoeck96}. In Table~\ref{tb:msb} we
give the values for $\delta^{12}_{pt}$ in the
$\overline{\mbox{MS}}$ scheme, in NLO and NNLO, as a function of
$\ln(m_{\tau}/\Lambda^{(3)}_{\overline{MS}})$. For completeness,
we also include precise values for $\delta^{00}_{pt}$, which may
be compared with those given previously in \cite{bra92}. Fitting the
experimental results for $R^{00}_{\tau,V}$ and $D^{12}_{\tau,V}$
we obtain in NNLO
$\Lambda^{(3)}_{\overline{MS}}=441\pm32\,\mbox{MeV}$, which
corresponds to $\alpha_{s}(m_{\tau}^{2})=0.367\pm0.018$ and
$\alpha_{s}(m_{Z}^{2})=0.1238\pm0.0018$. This is very close to
the value $\alpha_{s}(m_{\tau}^{2})=0.360\pm0.022$ obtained by
ALEPH in a fit involving more $R^{kl}_{\tau}$ moments and
$D=4,6,8$ nonperturbative contributions \cite{hoeck96}. We also obtain
$\delta^{00}_{(6)V}=0.0147\pm0.0025$, which is in reasonable
agreement with the value of the nonperturbative contribution
obtained by ALEPH \cite{hoeck96}. This confirms our expectation that
the $D=6$ approximation adopted in this work provides a good
approximation to the more complete analysis involving larger set
of parameters.

\begin{table}
\begin{center}
\begin{tabular}{||c||c|c||c|c||}
\hline
$\ln(m_{\tau}/\Lambda^{(3)}_{\overline{MS}})$ & 
$\delta_{\overline{MS},NL}^{00}$ & 
$\delta_{\overline{MS},NNL}^{00}$ &
$\delta_{\overline{MS},NL}^{12}$ & 
$\delta_{\overline{MS},NNL}^{12}$ \\
\hline
1.30 &    0.1967 &   0.2285 &   0.1267 &    0.1420  \\

1.35 &    0.1891  &  0.2187 &   0.1235 &    0.1364 \\

1.40 &    0.1820  &  0.2095  &  0.1206 &    0.1315 \\

1.45   &  0.1753  &  0.2009 &   0.1178 &    0.1273  \\

1.50  &   0.1690  &  0.1928 &   0.1151 &    0.1235 \\

1.55  &   0.1631 &   0.1852 &   0.1126 &    0.1201 \\

1.60  &   0.1576 &   0.1781 &   0.1102 &    0.1170 \\

1.65  &   0.1523 &   0.1714  &  0.1078 &    0.1142 \\

1.70   &  0.1474  &  0.1652 &   0.1056  &   0.1115 \\

1.75  &   0.1428  &  0.1593 &   0.1034 &    0.1090 \\

1.80   &  0.1384 &   0.1538 &   0.1013  &   0.1066 \\

1.85  &   0.1342  &  0.1486  &  0.0993  &   0.1043 \\

1.90   &  0.1303 &   0.1437 &   0.0973  &   0.1021 \\
\hline
\end{tabular}   
\end{center}
\caption{Table of values of the NLO and NNLO predictions for
$\delta^{00}_{pt}$ and $\delta^{12}_{pt}$ in the 
$\overline{\mbox{MS}}$ scheme.}
\label{tb:msb}
\end{table}

Let us now consider the same fit, but in a different
renormalization scheme. As is well known, the theoretical and
phenomenological motivation for the widely used
$\overline{\mbox{MS}}$ scheme is not very strong, and there has
been extensive discussion on the problem of the optimal choice of
the renormalization scheme \cite{stev81,dhar83a,dhar83b,grun84},
\cite{grun80}--\cite{duke85}. One of the interesting 
approaches is based on the so called principle of minimal
sensitivity (PMS) \cite{stev81}. The philosophy behind this approach
is very simple --- since the theoretical predictions of any
theory should be in principle independent of the RS, then in the
finite order of perturbation expansion one should look for the
RS, which mimics this as close as possible.

In the case of the conventional perturbative QCD expansion the RS
parameters of the PMS scheme are determined by a system of
transcendental and algebraic equations \cite{stev81}. Unfortunately,
in the case of perturbative predictions obtained via numerical
evaluation of the contour integral these equations do not apply,
so the optimized parameters have to be determined by direct
computation of $\delta^{kl}$ for differnt values of $r_{1}$ and
$c_{2}$.
 
The dependence of $\delta_{pt}^{00}$ on the scheme parameters
$r_{1}$ and $c_{2}$ was discussed in detail in \cite{racz96a} and the
RS dependence of $\delta_{pt}^{12}$ was investigated in
\cite{racz96b,racz96c}. In both cases it was found that for
moderate values of 
$\Lambda^{(3)}_{\overline{MS}}$ the NNLO predictions have a
saddle point type of behavior as a function of $r_{1}$ and
$c_{2}$ and that the position of the saddle point is well
approximated by $r_{1}=0$ and $c_{2}=1.5\rho_{2}=7.857$.
(Incidentally, these scheme parameters correspond to the
approximate solution \cite{penn82} of the algebraic PMS equations for
$\delta_{\Pi}$ evaluated for spacelike momenta.) For very large
values of $\Lambda^{(3)}_{\overline{MS}}$ the RS-dependence
pattern is more complicated than a simple saddle point, but even
then the scheme parameters distinguished above belong to the
region of extremely small RS dependence. We shall therefore
accept these parameters as the PMS parameters in NNLO. The values
of the NLO and NNLO PMS predictions for $\delta_{pt}^{12}$ are
given in Table~\ref{tb:pms}. (The values for $\delta_{pt}^{00}$
have been already given in \cite{racz96a}. The contour plots provided
there in principle allow one to obtain predictions for
$\delta_{pt}^{00}$ in arbitrary scheme with reasonably large
expansion coefficients.)

\begin{table}
\begin{center}
\begin{tabular}{||c||c|c||}
\hline
$\ln(m_{\tau}/\Lambda^{(3)}_{\overline{MS}})$ & 
$\delta_{PMS,NL}^{12}$ & 
$\delta_{PMS,NNL}^{12}$ \\
\hline

1.30 &    0.1382 &  0.1634 \\

1.35 &   0.1329 &  0.1521\\

1.40 &   0.1284 &  0.1432\\

1.45 &   0.1245 &  0.1360 \\

1.50 &   0.1209 &  0.1300\\

1.55 &   0.1177 &  0.1250\\

1.60  &  0.1148 &  0.1208\\

1.65  &  0.1120  &  0.1171\\

1.70  &  0.1095 &  0.1138\\

1.75  &  0.1070 &   0.1108\\

1.80  &  0.1047 &  0.1081\\

1.85  &  0.1025 &  0.1055\\

1.90   & 0.1004 &  0.1032\\
\hline
\end{tabular}   
\end{center}
\caption{Table of values of the NLO and NNLO predictions for
$\delta^{12}_{pt}$ obtained in a scheme preferred by the
principle of minimal sensitivity ($r_{1}=-0.64$ in NLO and
$r_{1}=0$, $c_{2}=1.5\rho_{2}=7.857$ in NNLO).}
\label{tb:pms}
\end{table}

Using the PMS predictions we obtain from the NNLO fit in the
vector channel $\delta^{00}_{(6)V}=0.0156\pm0.0023$ and
$\Lambda^{(3)}_{\overline{MS}}=421\pm30\,\mbox{MeV}$, which
corresponds to $\alpha_{s}(m_{\tau}^{2})=0.356\pm0.017$ and
$\alpha_{s}(m_{Z}^{2})=0.1226\pm0.0018$. We see that in NNLO the
change from the $\overline{\mbox{MS}}$ scheme to the PMS scheme
results in the reduction of the fitted value of $\alpha_{s}$ by
an amount significant compared for example to the presently
available experimental precision.
 
\begin{figure}[htb]
 ~\epsfig{file=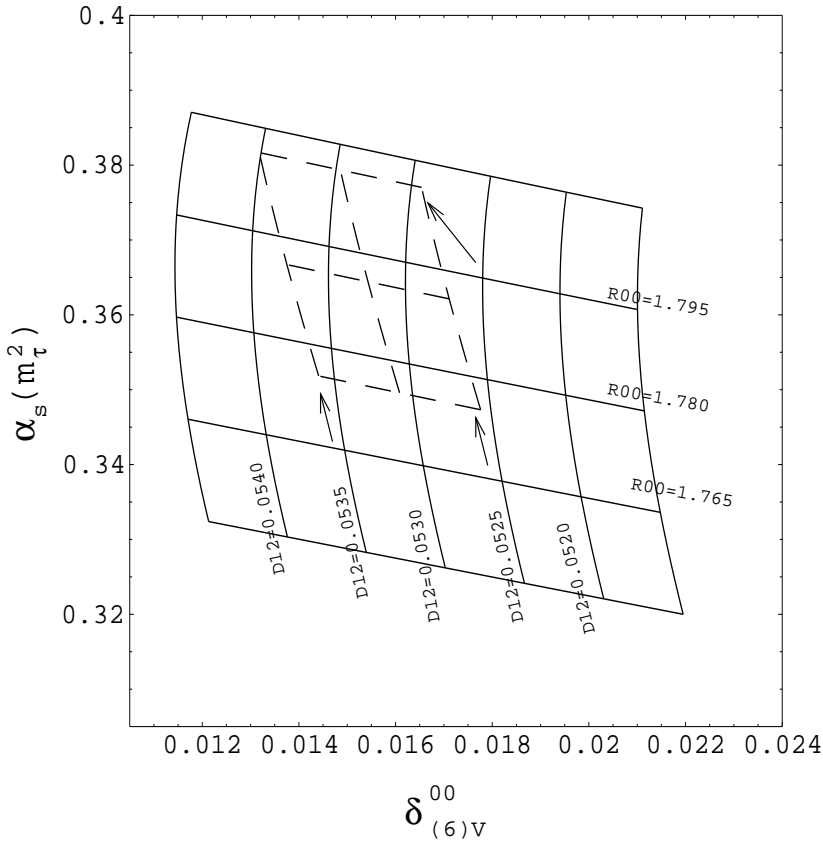,bbllx=0pt,bblly=0pt,bburx=308pt,bbury=318pt} 
\caption{Plot of the fitted values of $\alpha_{s}(m^{2}_{\tau})$
and $\delta^{00}_{(6)}$ in the vector channel as a function of
$R^{00}_{\tau,V}$ and $D^{12}_{\tau,V}$, obtained using the NNLO
PMS predictions. The dashed lines indicate the change in the plot
when the $\overline{\mbox{MS}}$ NNLO predictions are used
instead.}
\label{fig:ald6vec}
\end{figure}

In order to make our calculations  more generally useful  we show in
Fig.~\ref{fig:ald6vec} the results of the NNLO fit of
$\alpha_{s}(m^{2}_{\tau})$ and 
$\delta^{00}_{(6)V}$, obtained using the PMS predictions, 
as a  function of the experimental values of $R^{00}_{\tau,V}$ and
$D^{12}_{\tau,V}$. For
comparison we indicate the results of the NNLO fit in the
$\overline{\mbox{MS}}$ scheme (dashed lines).

Besides PMS another frequently discussed approach to the
optimization is the effective charge (EC) method \cite{grun80,grun84}, which
amounts to the absorption of all the higher order radiative
corrections to the physical quantity --- in this case
$\delta_{\Pi}$ --- into the definition of the coupling constant.
In the EC scheme we have $r_{1}=0$ and 
$c_{2}=\rho_{2}$. The NLO and NNLO predictions for
$\delta_{pt}^{00}$ and $\delta_{pt}^{12}$ in the EC scheme are
given in Table~\ref{tb:ec}. We see that in NNLO the difference
between EC and PMS is very small. This is reflected by the
results of the fit: in NNLO we obtain
$\alpha_{s}(m_{\tau}^{2})=0.356$ and $\delta^{00}_{(6)V}=0.0158$.

\begin{table}
\begin{center}
\begin{tabular}{||c||c|c||c|c||}
\hline
$\ln(m_{\tau}/\Lambda^{(3)}_{\overline{MS}})$& 
$\delta_{EC,NL}^{00}$ & 
$\delta_{EC,NNL}^{00}$ &
$\delta_{EC,NL}^{12}$ & 
$\delta_{EC,NNL}^{12}$ \\
\hline
                    1.30 &    0.2151&   0.2357 &  0.1360 &   0.1624\\

                    1.35  &  0.2063 &  0.2261 &  0.1312 &  0.1524\\

                    1.40  &  0.1979 &  0.2168 &  0.1271 &  0.1439\\

                    1.45 &   0.1901 &  0.2080 &   0.1233 &  0.1368\\

                    1.50 &   0.1828 &  0.1996 &  0.1200  &   0.1308\\

                    1.55 &   0.1759 &  0.1917 &  0.1169 &  0.1257\\

                    1.60 &   0.1694 &  0.1842 &  0.1141 &  0.1213\\

                    1.65  &  0.1634  & 0.1771 &  0.1114 &  0.1175\\

                    1.70 &   0.1577 &  0.1704 &  0.1089 &  0.1141\\

                    1.75 &   0.1523 &  0.1642 &  0.1065 &  0.1110\\

                    1.80 &   0.1473 &  0.1583 &  0.1043 &  0.1082\\

                    1.85 &   0.1425 &  0.1528 &  0.1021 &  0.1057\\

                    1.90 &   0.1380 &  0.1476 &  0.1000 &  0.1033\\
\hline
\end{tabular}   
\end{center}
\caption{Table of values of the NLO and NNLO predictions for
$\delta^{00}_{pt}$ and 
$\delta^{12}_{pt}$ in the Effective Charge scheme.  }
\label{tb:ec}
\end{table}

To have a broader picture of the renormalization scheme
dependence we perform the fit in a more general class of schemes.
It is clear that regardless of our choice of the concrete optimal
scheme there is a continuum of schemes, which seem to be equally
reasonable. Predictions in such schemes should also be somehow
taken into account in the phenomenological analysis. A natural
way to do this is to supplement the prediction in a preferred
scheme with an estimate of the variation of the predictions over
a whole set of {\em a priori} acceptable schemes. A concrete
realization of this idea was presented in \cite{racz95}, based on the
existence of the RS invariant $\rho_{2}$, which provides a
natural RS independent characterization of the magnitude of the
NNLO corrections for the considered physical quantity. In
\cite{racz95} it was proposed to calculate variation of the
predictions over the set of schemes for which the expansion
coefficients satisfy the condition:
\begin{equation}
\sigma_{2}(r_{1},r_{2},c_{2}) \leq l\,|\rho_{2}|,
\label{eq:condition}
\end{equation}
 where  
\begin{equation}
\sigma_{2}(r_{1},r_{2},c_{2})=|c_{2}|+|r_{2}|+c_{1}|r_{1}|+r_{1}^{2}.
\end{equation}
A motivation for the condition (\ref{eq:condition}) is that it
 eliminates schemes in which the expansions (\ref{eq:dpert}) and
 (\ref{eq:rge}) involve unnaturally large expansion coefficients that
 introduce large cancellations in the expression for the RS
 invariant $\rho_{2}$. The constant $l\,$ in the condition
 (\ref{eq:condition}) controls the degree of cancellation that we
 want to allow in the expression for $\rho_{2}$. In our case we
 have for the PMS parameters $\sigma_{2}(\mbox{PMS})\approx2|\rho_{2}|$,
 so in order to take into account the schemes, which have the
 same --- or smaller --- degree of cancellation as the PMS scheme
 we take $l=2$. One may expect, that the estimate of the RS
 dependence obtained according to this prescription would be
 useful for a quantitative comparison of reliability of
 perturbative predictions for different physical quantities,
 evaluated at different energies. It is also clear that any large
 variation of the predictions over a set of schemes satisfying
 the constraint (\ref{eq:condition}) with $l=2$ would be an
 unambiguous sign of a limited applicability of the NNLO
 expression.

\begin{figure}[htb]
 ~\epsfig{file=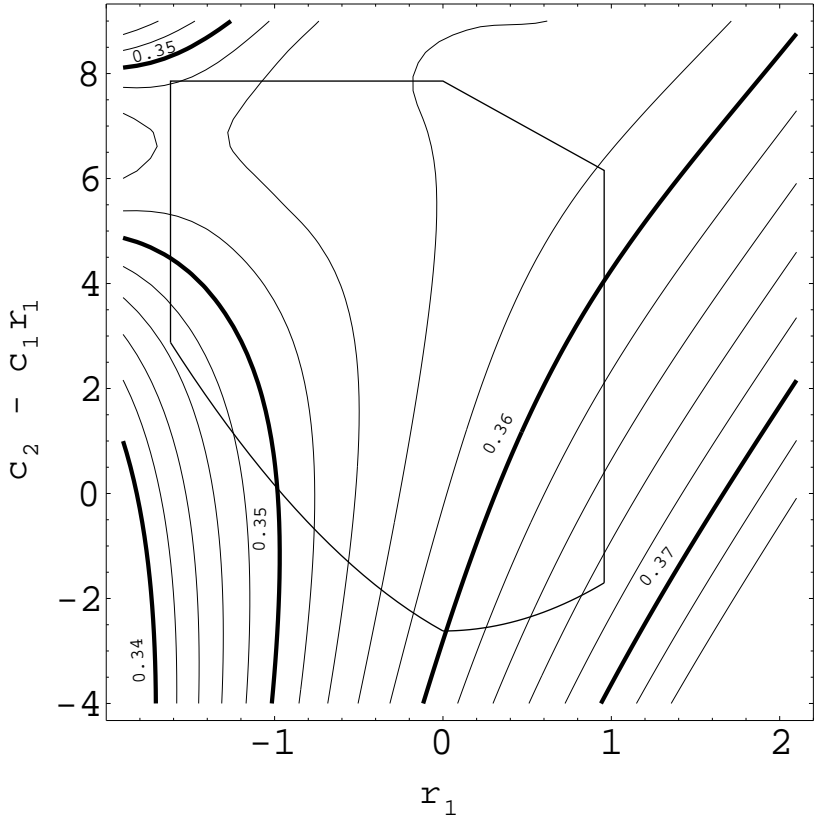,bbllx=0pt,bblly=0pt,bburx=303pt,bbury=309pt} 
\caption{The contour plot of the fitted value of
$\alpha_{s}(m^{2}_{\tau})$ in the vector channel as a function of
the RS parameters $r_{1}$ and $c_{2}$. For technical reasons we
use $c_{2}-c_{1}r_{1}$ as an independent variable on the vertical
axis. The region of the scheme
parameters satisfying the condition
(\protect{\ref{eq:condition}}) with $l=2$ has 
been also indicated.}
\label{fig:alfitvecrsdep}
\end{figure}

In Fig.~\ref{fig:alfitvecrsdep} we show how the value of
$\alpha_{s}(m_{\tau}^{2})$ resulting from the fit depends on the
parameters $r_{1}$ and $c_{2}$ specifying the renormalization
scheme in NNLO. In the region of scheme parameters satisfying
condition~(\ref{eq:condition}) with $l=2$ the minimum value of
$\alpha_{s}(m^{2}_{\tau})=0.347$
($\Lambda^{(3)}_{\overline{MS}}=403\,\mbox{MeV}$,
$\alpha_{s}(m^{2}_{Z})=0.1217$) is attained for $r_{1}=-1.62$ and
$c_{2}=0$, and the maximum value of
$\alpha_{s}(m^{2}_{\tau})=0.367$
($\Lambda^{(3)}_{\overline{MS}}=440\,\mbox{MeV}$,
$\alpha_{s}(m^{2}_{Z})=0.1238$) is attained for $r_{1}=0.96$ and
$c_{2}=0$. It should be noted that the $\overline{\mbox{MS}}$
scheme parameters lie outside the $l=2$ region, but the fitted
value of $\alpha_{s}(m^{2}_{\tau})$ in this scheme coincides with
the maximal value quoted above. It is also important to note that
one of the other potentially interesting schemes, for which the
NNLO expressions recently became available \cite{peter97} --- the so
called V~scheme \cite{fisch77} --- corresponds to the RS parameters
$r_{1}=-0.109$ and $c_{2}=26.200$, which lie very far outside the
$l=2$ region. It seems that the large value of $c_{2}$ in the
V~scheme restricts its usefulness for the low energy perturbative
predictions, because of the influence of the Landau pole in the
RG equation.

\begin{figure}[htb]
 ~\epsfig{file=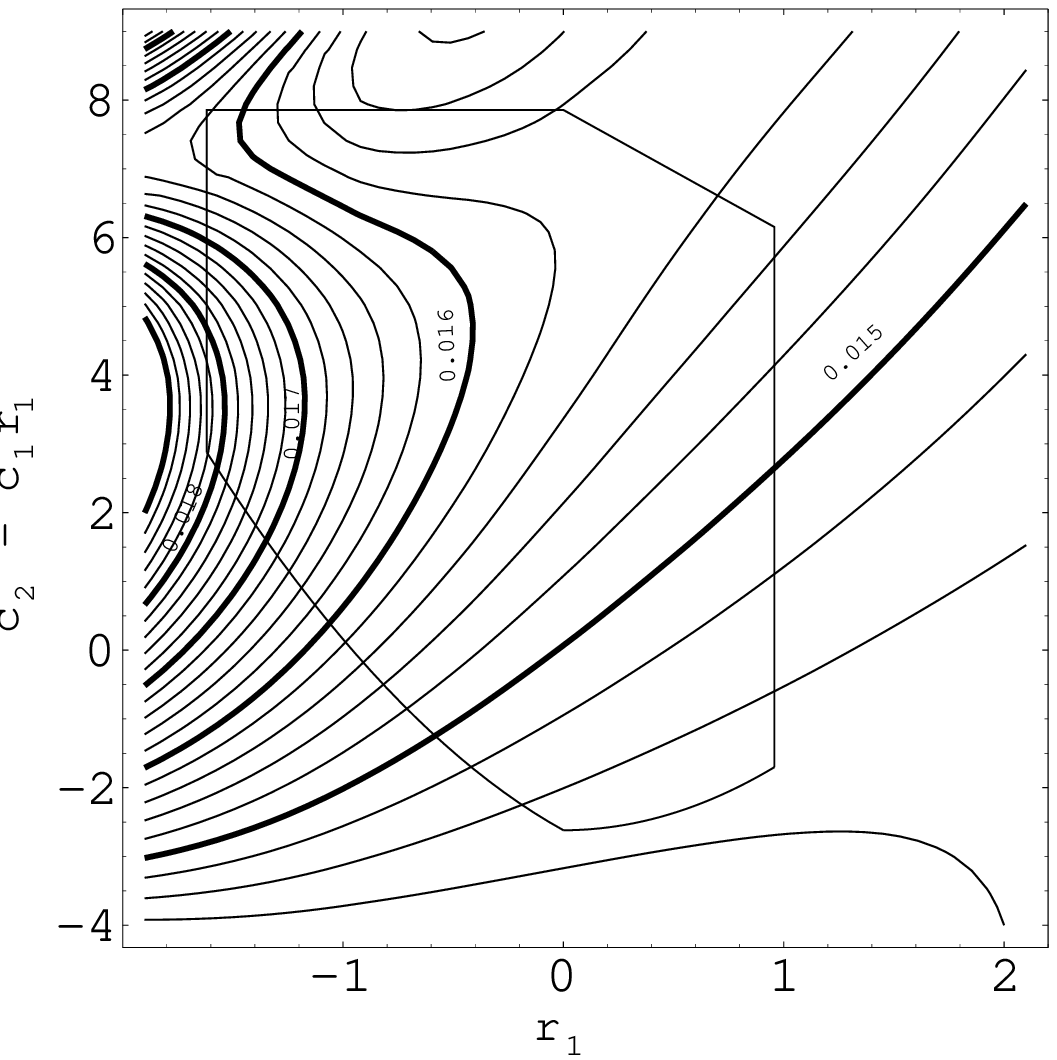,bbllx=0pt,bblly=0pt,bburx=303pt,bbury=304pt} 
\caption{The contour plot of the fitted value of
$\delta^{00}_{(6)}$ in the vector channel as a function of the
RS parameters $r_{1}$ and $c_{2}$. The region of the scheme
parameters satisfying the condition
(\protect{\ref{eq:condition}}) with $l=2$ has 
been also indicated.}
\label{fig:d6fitvecrsdep}
\end{figure}

It is of some interest to perform the same fits using instead the
NLO predictions. Using the NLO predictions in the
$\overline{\mbox{MS}}$ scheme, we obtain
$\delta^{00}_{(6)V}=0.0148$ and
$\Lambda^{(3)}_{\overline{MS}}=527\,\mbox{MeV}$, which translates
via NLO RG equation into $\alpha_{s}(m_{\tau}^{2})=0.388$, which
in turn corresponds via NLO extrapolation to
$\alpha_{s}(m_{Z}^{2})=0.1261$. We see that the value of
$\delta^{00}_{(6)V}$ 
is practically identical to that obtained in the NNLO fit.
However, the value of the strong coupling constant is 
surprisingly high and it is significantly different from the
NNLO value.
Using the NLO predictions in the PMS scheme we obtain
$\delta^{00}_{(6)V}=0.0150$ and
$\Lambda^{(3)}_{\overline{MS}}=465\,\mbox{MeV}$, which
corresponds to $\alpha_{s}(m_{\tau}^{2})=0.358$ and
$\alpha_{s}(m_{Z}^{2})=0.1230$. We see that although the
difference in the NLO and NNLO values of
$\Lambda^{(3)}_{\overline{MS}}$ obtained in the PMS fits is still
considerable, it is nevertheless much smaller than in the case of
the $\overline{\mbox{MS}}$ scheme. (If we look at
$\alpha_{s}(m_{\tau})$ instead of $\Lambda^{(3)}_{\overline{MS}}$
this difference is even smaller, although this may be partially a
result of a fortuitous compensation between different factors
influencing the evaluation of $\alpha_{s}$ in
various orders.) For completeness we give the NLO results in the
EC scheme: $\delta^{00}_{(6)V}=0.0149$ and
$\Lambda^{(3)}_{\overline{MS}}=472\,\mbox{MeV}$
($\alpha_{s}(m_{\tau}^{2})=0.361$). These numbers provide a nice
illustration of the fact, that the difference between the NLO and
NNLO predictions is strongly scheme dependent. Therefore, if such
a difference is to be used in any way to estimate the precision
of the QCD prediction, some way of making a preferred choice of
the renormalization scheme must be employed.

So far our discussion was concentrated primarily on the value of
$\alpha_{s}(m_{\tau}^{2})$ coming from the fit. In
Fig.~\ref{fig:d6fitvecrsdep} we show the RS dependence of the fitted
value of $\delta^{00}_{(6)V}$. We find that in the set of schemes
satisfying the constraint (\ref{eq:condition}) with $l=2$ the
fitted value of $\delta^{00}_{(6)V}$ changes in the range
0.0145--0.0182. Using the Eq.~(\ref{eq:vacsat}) we may translate this
into the range $(2.22-2.78)\times10^{-4}\,\mbox{GeV}^{6}$ for the
commonly used parameter $\alpha_{s}\rho<\bar{q}q>^{2}$. This
seems to be in reasonable agreement with the values obtained
previously by other authors, for example
$1.8\times10^{-4}\mbox{GeV}^{6}$ obtained in the original work of
SVZ \cite{shif79} or $(3.8\pm2.0)\times10^{-4}\mbox{GeV}^{6}$
obtained in a more recent analysis \cite{gime91}.

\section{Fits in the axial-vector channel}

Similarly as in the case of the vector channel we start with the
fit in the $\overline{\mbox{MS}}$ scheme. We obtain
$\delta^{00}_{(6)A}=-0.0168\pm0.0021$ and
$\Lambda^{(3)}_{\overline{MS}}=398\pm37\,\mbox{MeV}$, which
corresponds to $\alpha_{s}(m_{\tau}^{2})=0.344\pm0.019$ and
$\alpha_{s}(m_{Z}^{2})=0.1213\pm0.0021$. The result of this fit
should be compared with $\alpha_{s}(m_{\tau}^{2})=0.365\pm0.025$,
obtained by ALEPH in a more complete fit \cite{hoeck96}. We find that
the difference between our results and the ALEPH result is
slightly bigger than in the case of vector channel. It should be
noted however that the ALEPH fit in the axial-vector channel has
surprisingly large $D=4$ contribution, which would be difficult
to justify theoretically.

Performing the NNLO fit in the PMS scheme we obtain
$\delta^{00}_{(6)A}=-0.0165\pm0.0018$ and
$\Lambda^{(3)}_{\overline{MS}}=380\pm34\,\mbox{MeV}$, which
corresponds to $\alpha_{s}(m_{\tau}^{2})=0.335\pm0.018$ and
$\alpha_{s}(m_{Z}^{2})=0.1203\pm0.0021$. Similarly as in the case
of the vector channel we find appreciable reduction in the
extracted value of $\alpha_{s}(m^{2}_{\tau})$ as compared to the
values obtained in the $\overline{\mbox{MS}}$ scheme.

\begin{figure}[htb]
 ~\epsfig{file=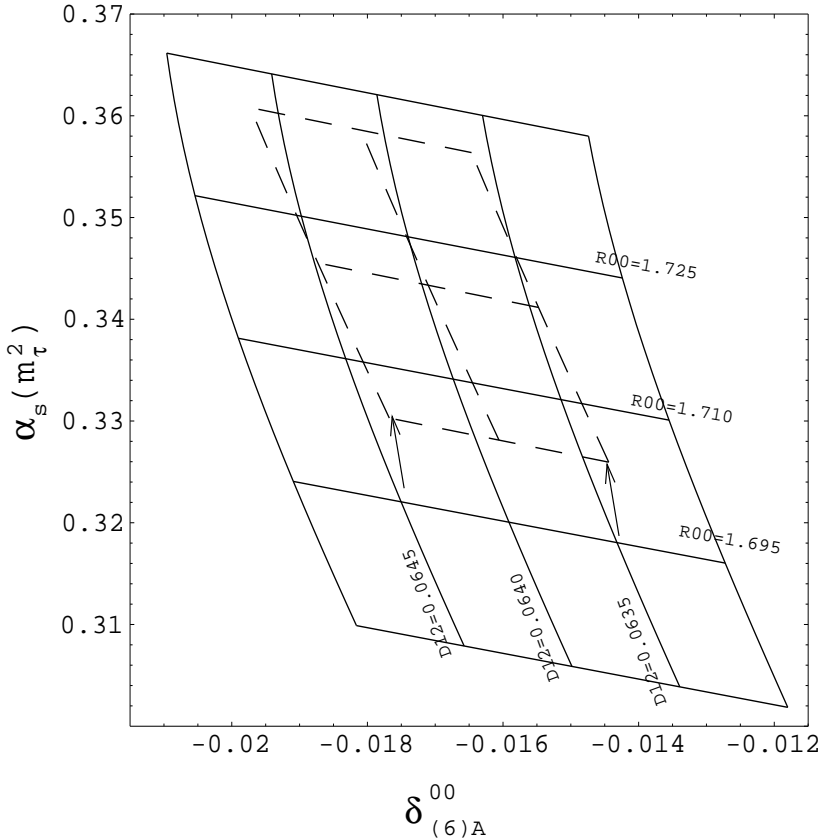,bbllx=0pt,bblly=0pt,bburx=308pt,bbury=318pt} 
\caption{Plot of the fitted values of $\alpha_{s}(m^{2}_{\tau})$
and $\delta^{00}_{(6)}$ in the axial-vector channel as a
function of $R^{00}_{\tau,A}$ and $D^{12}_{\tau,A}$, obtained
using the NNLO PMS predictions. The dashed lines indicate the
change in the plot when the $\overline{\mbox{MS}}$ NNLO
predictions are used instead.}
\label{fig:ald6ax}
\end{figure}

To make our results useful in the case of future improvements of
the experimental analysis we show in Fig.~\ref{fig:ald6ax} the
results of the NNLO fit of $\alpha_{s}(m^{2}_{\tau})$ and
$\delta^{00}_{(6)A}$, obtained with the PMS predictions, as a
function of the experimental values of $R^{00}_{\tau,A}$ and
$D^{12}_{\tau,A}$. For comparison we indicate the results of the
NNLO fit in the $\overline{\mbox{MS}}$ scheme (dashed lines).

\begin{figure}[htb]
 ~\epsfig{file=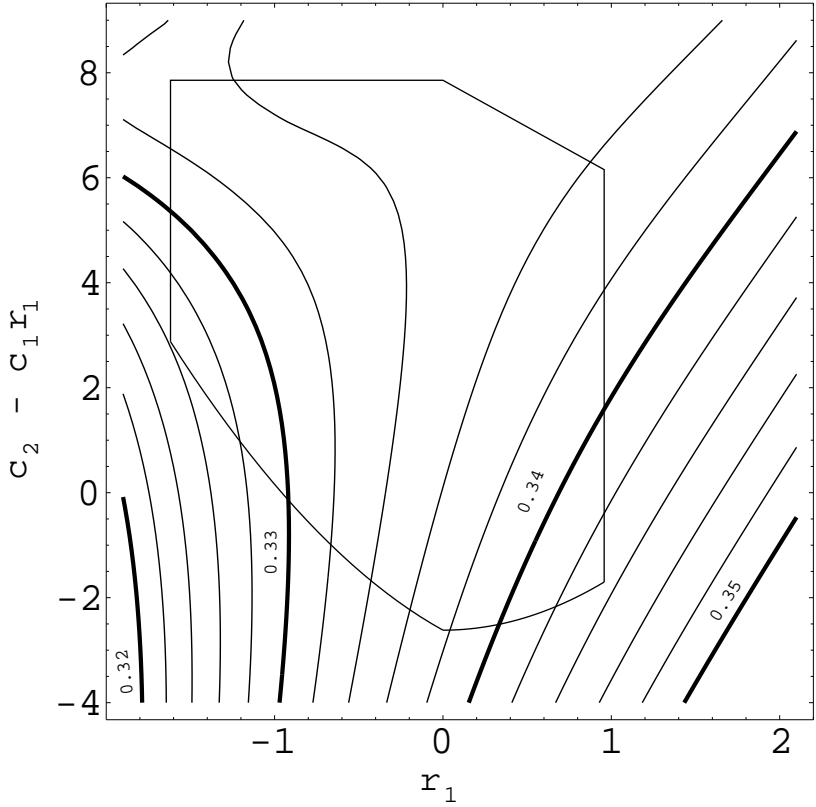,bbllx=0pt,bblly=0pt,bburx=303pt,bbury=304pt} 
\caption{The contour plot of the fitted value of
$\alpha_{s}(m^{2}_{\tau})$ in the axial-vector channel as a
function of the RS parameters $r_{1}$ and $c_{2}$. The region of
the scheme parameters satisfying the condition
(\protect{\ref{eq:condition}}) with 
$l=2$ has been also indicated.}
\label{fig:alfitaxrsdep}
\end{figure}

In Fig.~\ref{fig:alfitaxrsdep} we show how the fitted value of
$\alpha_{s}(m^{2}_{\tau})$ depends on the RS parameters $r_{1}$
and $c_{2}$. In the region of the 
scheme parameters satisfying the condition~(\ref{eq:condition}) with $l=2$ we
have $0.326<\alpha_{s}(m^{2}_{\tau})<0.343$
($364\,\mbox{MeV}<\Lambda^{(3)}_{\overline{MS}}<397\,\mbox{MeV}$,
$0.1193<\alpha_{s}(m^{2}_{Z})<0.1212$). Similar figure may be
obtained for the RS dependence of $\delta^{00}_{(6)A}$: it is found that
$0.015<-\delta^{00}_{(6)A}<0.017$.

Having obtained the results for $\delta^{00}_{(6)}$ in the vector
and axial-vector channels it is of some interest to verify the
simple relation between them, implied by the generalized vacuum
saturation approximation:
$\delta^{00}_{(6)A}/\delta^{00}_{(6)V}=-11/7\approx-1.57$. Using
the numbers from the NNLO PMS fits in both channels we obtain
$\delta^{00}_{(6)A}/\delta^{00}_{(6)V}=-1.06\pm0.28$.
 
In order to have a full picture of the perturbative uncertainties
in the axial-vector channel we also perform the NLO fits. Using
the $\overline{\mbox{MS}}$ scheme, we obtain
$\delta^{00}_{(6)A}=-0.0166$ and
$\Lambda^{(3)}_{\overline{MS}}=473\,\mbox{MeV}$, which translates
via NLO RG equation into $\alpha_{s}(m_{\tau}^{2})=0.362$, which
in turn corresponds via NLO extrapolation to
$\alpha_{s}(m_{Z}^{2})=0.1235$. In the PMS scheme we obtain
$\delta^{00}_{(6)A}=-0.0167$ and
$\Lambda^{(3)}_{\overline{MS}}=419\,\mbox{MeV}$, which implies
$\alpha_{s}(m_{\tau}^{2})=0.336$ and
$\alpha_{s}(m_{Z}^{2})=0.1207$. Similarly as in the vector
channel case we note a rather large difference between NLO and
NNLO in the case of the $\overline{\mbox{MS}}$ scheme, which is
significantly reduced if the preferred scheme is PMS. In NLO in
the EC scheme we obtain $\delta^{00}_{(6)A}=-0.0167$ and
$\Lambda^{(3)}_{\overline{MS}}=425\,\mbox{MeV}$
($\alpha_{s}(m_{\tau}^{2})=0.339$).

\section{Discussion and conclusions}

Our discussion of the RS dependence of the QCD predictions and
the fits may be summarized as follows:

\begin{itemize}

\item[1.] Changing the scheme from the  $\overline{\mbox{MS}}$
scheme to the PMS scheme we obtain a reduction in the extracted
value of $\alpha_{s}(m_{\tau}^{2})$ by approximately 0.01
($\alpha_{s}(m_{Z}^{2})$ is reduced by 0.001). Also, the
difference between the NLO and NNLO results is much smaller in
the PMS scheme than in the $\overline{\mbox{MS}}$ scheme. 

\item[2.] Varying the scheme parameters $r_{1}$ and $c_{2}$ in
the region satisfying the condition~(\ref{eq:condition}) with $l=2$ we
obtain an uncertainty in the extracted value of $\alpha_{s}(m_{\tau}^{2})$
of approximately 0.02 (uncertainty in $\alpha_{s}(m_{Z}^{2})$ is 
0.002). As was argued in the text, this set of scheme parameters
seems to be a minimal set that one should take into account in
the discussion of the RS dependence. 

\end{itemize}

Our general conclusion is that the perturbative predictions for
the QCD effects in the inclusive decay rates for the semileptonic
$\tau$ decays appear to be relatively precise, despite the rather
low energy scale. It should be emphasized however, that in the
discussion of the final precision of $\alpha_{s}$ extracted from
the $\tau$ decays one should also take into account the
approximate character of the SVZ expansion itself
\cite{shif79,nari95b,chibi97}.

Discussions with R. Alemany, M. Beneke, M. Davier, A. H\"{o}cker,
C. Maxwell and M. Neubert are gratefully acknowledged.

\end{document}